\documentclass[twocolumn,10pt,twoside]{IEEEtran}
\usepackage{epsfig}
\usepackage{amsmath}
\usepackage{theorem}
\usepackage{amsmath}
\usepackage{cancel}
\usepackage{mathrsfs}
\usepackage{amsfonts}
\usepackage{amssymb}
\usepackage{mathrsfs}
\usepackage{euscript}
\usepackage{graphicx}
\usepackage{multirow}
\usepackage{algorithm}
\usepackage{algorithmic}
\usepackage{rotating}
\usepackage{subfigure}
\usepackage{cite}
\usepackage{url}
\usepackage{color}
\usepackage{amsmath}
\usepackage{lipsum}

\IEEEoverridecommandlockouts

\newtheorem{definition}{Definition}

\begin{document}

	\title{{\huge Multi-objective Resource Allocation for D2D and Enabled MC-NOMA Networks by Tchebycheff Method} }
	\vspace{-10mm}
	\author{\IEEEauthorblockN{Siavash Bayat, \textit{Member, IEEE}, Ata Khalili, \textit{Member, IEEE}, Shayan Zargari, Mohammad Robat Mili, \textit{Member, IEEE}, and Zhu Han,~\textit{Fellow, IEEE}}
		\vspace{-10mm}
		\thanks{Siavash Bayat, Ata Khalili, Shayan Zargari, and Mohammad Robat Mili are with the Electronics Research Institute, Sharif University of Technology, Tehran, Iran. (e-mails: bayat@sharif.edu, khalili.ata@sharif.edu.~shayanzargari66@gmail.com.~mohammad.robatmili@ieee.org).~Zhu Han is with the Electrical and Computer Engineering Department, University of Houston, TX,~77004,
			USA. (e-mail:~hanzhu22@gmail.com).}}
	\maketitle
	\vspace{-10mm}
	\begin{abstract}
		This paper considers a resource allocation problem in device-to-device (D2D) communications sharing the same frequency spectrum.~In particular,~the cellular users (CUs) utilize non-orthogonal multiple access (NOMA) while D2D users (DUs) adopt the orthogonal frequency division multiple access (OFDMA).~\textcolor{blue}{A multi-objective optimization problem (MOOP) is formulated, which jointly maximizes the sum rate of DUs and CUs in uplink communications while taking into account the maximum transmit power budget and minimum data rate requirement for DUs and CUs.} This MOOP is handled by the weighted Tchebycheff method, which converts it into a single-objective optimization (SOOP). Then, the monotonic optimization approach is employed to solve this SOOP optimally. Numerical results unveil an interesting tradeoff between D2D and CUs.
	\end{abstract}
	\vspace{-8mm}
	\section{Introduction}
	\vspace{-1mm}
	Device-to-Device (D2D) communication is promoted as an innovative paradigm to improve the network performance as well as the system resources utilization in the fifth-generation (5G) cellular networks and beyond. The power and channel allocation for D2D communication need elaborate coordination with cellular users (CUs), as D2D users (DUs) can impose interference to other users \cite{new1}. 
	In practice, D2D communication operates either in the overlay or underlay modes along with existing CUs. In fact, the underlay mode is appealing as it follows the system to achieve higher spectral efficiency in which the spectrum is shared between DUs and CUs\cite{def5}.
	
	Non-orthogonal multiple access (NOMA) has been proposed as one of the fundamental techniques for beyond 5G to achieve a better balance between system spectral efficiency (SE) and user fairness \cite{new2}.~As a result, the integration between NOMA and D2D communication has significant attention in order to improve user connectivity as well as SE \cite{Single2,5}.~In this regard, resource allocation is one challenging problem which can appropriately mitigate interference, thereby improving the system SE.~In \cite{Single2}, while using the successive interference cancellation (SIC) to detect the multiplexing signals, the resource block assignment and power allocation were optimized to maximize the sum data rate of the D2D pairs. In \cite{5}, according to interference status, different NOMA-aided spectrum-sharing modes (i.e., the D2D access scheme) for the paired DUs and CUs were considered. Then, a connectivity-maximization problem was formulated under mode selection, user pairing, and power control while guaranteeing the decoding thresholds of CUs and DUs to take advantage of the NOMA-and-D2D integrated structure. In \cite{new3}, the authors offered an innovative resource allocation policy to enhance the performance of D2D communications underlying CUs in the downlink (DL).
	For a network consisting of CUs and DUs in \cite{R1}, the problem of power allocation and user clustering was studied while the sum-rate of the NOMA-based network was maximized. In \cite{R2}, a joint power allocation and user scheduling for the D2D-enabled HetNets with NOMA was proposed to maximize the ergodic sum rate of the near users in the small cells, while guaranteeing the quality-of-service requirements of the far users and the macro-cell users. The authors in \cite{R3} offered a joint optimization framework for D2D-enabled NOMA networks, where the performance of the D2D communication was maximized while considering the SIC decoding order of the NOMA-based CU equipment. In \cite{R4}, the mode selection and resource allocation problem for D2D-enabled NOMA cellular networks was considered. However, the inter-lay mode was introduced for D2D communications in NOMA system to maximize the system sum rate, which exploited the SIC to cancel the interference between D2D pairs and CUs. In \cite{R5}, a NOMA-enhanced D2D communication scheme was considered and then the system sum-rate was maximized, while optimizing subchannel and power allocation. Then, a solution was proposed to assign subchannels to D2D groups and allocate power to receivers in each D2D group. The authors in \cite{R6} considered the resource allocation problem for an uplink multi-carrier NOMA in D2D underlaid cellular networks and then the maximization problem of the Nash product of each user was investigated as a Nash bargaining game.
	

\textcolor{blue}{Despite the fruitful results in the literature, the performance of the D2D communications can still be improved as the conflicting goals of D2D and cellular create  serious network performance bottlenecks.~In the literature, the framework of the multi-objective optimization problem (MOOP) was employed to address the conflicting objectives in wireless systems [13]-[15].~Regarding this, the authors in [13] proposed a MOOP for maximizing the signal-to-interference-plus-noise ratio (SINR) to determine the optimal power allocation for each D2D pair. A MOOP trade-off was analyzed in [14] to investigate the trade-off between EE and SE in a D2D underlying system. This problem was converted into a SOOP via the $\epsilon$-method, and a two-stage iterative algorithm was propose.~However, a non-trivial trade-off between the DUs and the CUs would be expected.~As a result, the spectrum sharing deployment under such networks suffers from inter-cell interference originated from the DUs and CUs in each cell which leads to an exciting optimization problem.}

 \textcolor{blue}{Nonetheless to the best of the authors' knowledge the optimal subchannel assignment and power allocation were not discussed in [4]-[12]. In particular, designing the optimal resource management in D2D-enabled NOMA cellular networks in underlay mode is challenging. Additionally, deriving achievable rate regions for the DUs and CUs via employing the NOMA scheme leads to an interesting trade-off problem that is not investigated in [4]-[12]. The contributions of this letter are summarized as follows:}
	
	
	\begin{itemize}
	
	\item  In contrast to the existing literature e.g., [4]-[12], we study the performance trade-off between the DUs and the CUs in D2D networks underlying NOMA CUs to maximize the sum data rate of the DUs and CUs simultaneously.
	
	\item To study such a trade-off, we formulate a MOOP framework by jointly optimizing the transmit powers and subcarrier allocation policies that can be obtained by adjustable weighting parameters to execute the resource allocation policy.
	
	\item To solve the MOOP at hand, we first apply the weighted Tchebycheff method, which converts the MOOP into a SOOP. Then, a monotonic optimization method is proposed to obtain the optimal resource allocation policy.
	
	\item In the numerical results, we provide an interesting trade-off between CUs and DUs and also demonstrate the superiority of the MC-NOMA scheme as compared to multi-carrier orthogonal multiple access (MC-OMA) schemes. This also shows that our proposed scheme outperforms the proposed algorithm in [12].

\end{itemize}
	\vspace{-4mm}    
	\section{System\ Model}
		\vspace{-2mm} 
	In this paper, an uplink single-cell NOMA-based cellular network is considered which compromises of one BS to serve $M$ CUs and $K$ D2D links (DUs) given by $\mathcal{M}=\{1,...,M\}$ and $\mathcal{K}=\{1,...,K\} $, respectively.~The total system bandwidth of $B$ Hz is divided into a set of subchannels denoted by $\mathcal{N}=\{1,...,N\}$, which is shared between CUs and DUs so that each subchannel has bandwidth of $B_{c}=B/N$ Hz.~We define $h^{n}_{t,r}$ as the channel gain between transmitter $t$ and receiver $r$ over subcarrier $n$.~For simplify, the channel gain between the transmitter and BS is defined as $h^{n}_{t}$.~Denote the instantaneous channel power gains for the $k$-th DU to the BS as $h_{k}^{n}$ and the link between the $m$-th CU and the BS as $g_{m}^{n}.$ The transmitted power from the $k$-th DU over subchannel $n$ is expressed by $p_{k}^{n}$ and for the $m$-th CU is given by $\hat{p}_{m}^{n}$.~Furthermore,~the noise power spectral density is given by $N_{0}$.~It should be noted that the D2D devices are performed based on the OFDMA protocol due to closing to each other, which indicates that the desired signal is more considerable than the interference terms.~In this network, the BS can employ the SIC technique based on the descending order in the channel gain.
The channel
		gains between the users and the BS should satisfy $g_{m}>h_{k}>g_{i}$
		then the BS will decode the signals $x_{m}$, $x_{k}$, and $x_{i}$
		sequentially by using the SIC technique. On the contrary,
		since the signal strength of $x_{m}$
		at the D2D receiver $x_{k}$ is the strongest, the D2D receiver will directly decode its desired
		signal.~This constraint ensures that the BS can perform SIC properly.~As a result, the instantaneous SINR for each CU $m$ on subchannel $n$\ is given by
	\begin{equation}
		\textcolor{blue}{
	\gamma_{m,\text{Cellular}}^{n}=\frac{\psi _{m}^{n}\hat{p}_{m}^{n}|g_{m}^{n}|^{2}}{\sigma^{2}_{\text{CU}}+{\textstyle\sum\limits_{\substack{ i\in\mathcal{M} \backslash m \\g_{i}^{n}<g_{m}^{n}}}}\psi_{i}^{n}\hat{p}_{i}^{n}|g_{i}^{n}|^{2}+{\textstyle\sum\limits_{\substack{ j\in \mathcal{K} \\ h_{j}^{n}<g_{m}^{n}}}}\varphi _{j}^{n}p_{j}^{n}|h_{j}^{n}|^2}},  \label{m11}
	\end{equation}%
	where $\psi _{i}^{n}$ and $\varphi _{j}^{n}$ are the binary variables for subchannel allocation for the CU and DU, respectively.~If $\psi _{m}^{n}=1$, subchannel $n$ is allocated to the $m$-th CU and $\psi _{m}^{n}=0,$ otherwise. In a similar manner, if $\varphi
	_{j}^{n}=1$, subchannel $n$ is assigned to the $j$-th DU and ${\small \varphi_{j}^{n}=0},$ otherwise. In $(\ref{m11})$, the term ${\textstyle\sum_{\substack{(i\neq m)\in \mathcal{M},\; g_{i}^{n}<g_{m}^{n}}}}\psi_{i}^{n}\hat{p}_{i}^{n}|g_{i}^{n}|^2$ is the
	interference term from the CUs due to operating on the same subchannel based on the NOMA scheme, and the term ${\small {\textstyle\sum_{\substack{ j\in \mathcal{K},\;h_{j}^{n}<g_{m}^{n}}}}\varphi _{j}^{n}p_{j}^{n}|h
		_{j}^{n}|^2}$ corresponds to the interference from the DUs. In addition, the instantaneous received SINR at the $k$-th DU on subchannel $n$\ can be written as%
	\begin{equation}
	\textcolor{blue}{
		\gamma_{k,\text{D2D}}^{n}=\frac{\varphi
			_{k}^{n}p_{k}^{n}|h_{k}^{n}|^2}{\sigma^{2}_{\text{D2D}}+\sum%
			\limits_{m\in \mathcal{M}}\psi _{m}^{n}\hat{p}_{m}^{n}|g_{m,k}^{n}|^2+{%
				\textstyle\sum\limits_{j\in \mathcal{K}\backslash k}}\varphi_{j}^{n}p_{j}^{n}|h_{j,k}^{n}|^2}},
	\label{m1}
	\end{equation}%
where $g_{m,k}^{n}$ is the channel power gain between
		the $m$-th CU and the $k$-th DU receiver over subchannel $n$, and ${\textstyle}%
		\sum_{m\in \mathcal{M}}\psi _{m}^{n}\hat{p}_{m}^{n}|g_{m,k}^{n}|^2$ is the interference term arising from the CUs.~Besides, ${%
		\textstyle\sum_{{j\in \mathcal{K}\backslash k}}}\varphi
	_{j}^{n}p_{j}^{n}|h_{j,k}^{n}|^2$
	denotes the interference term
	resulting from the other D2D pairs, where $h_{j,k}^{n}$ is the instantaneous channel power gain from the $%
	j$-th DU transmitter to the $k$-th DU receiver on subchannel $n$.
		\vspace{-2mm}
	\section{Problem Formulation}
	\vspace{-2mm}
	In this section, to strike a tradeoff between the system performance of DUs and CUs, we formulate a MOOP in which jointly the sum rate of the DUs, $\text{R}_{\text{DU}}$, and CUs, $\text{R}_{\text{CU}}$, are maximized. The proposed MOOP framework aims to obtain power allocation as well as the subchannel assignment strategy to study the performance tradeoff between conflicting system objectives. This joint optimization can be formulated through the following MOOP: 
	\begin{subequations}
		\label{mm1}
		\textcolor{blue}{
			\begin{align}
			& \max_{\left\{ {\normalsize \mathbf{p}},\mathbf{\hat{p}},\boldsymbol{%
					\varphi },\boldsymbol{\psi }\right\} }\ \ \text{R}_{\text{DU}}=\sum_{k=1}^{K}
			\text{R}_{\text{DU,}k}  \label{m2} \\
			& \max_{\left\{ {\normalsize \mathbf{p}},\mathbf{\hat{p}},\boldsymbol{\varphi },\boldsymbol{\psi }\right\} }\text{ \ R}_{\text{CU}}=\sum_{m=1}^{M} \text{R}_{\text{CU,}m}  \label{m22} \\
			~~\mathrm{s.t.}~~& \sum_{n=1}^{N}\varphi _{k}^{n}p_{k}^{n} \leq p^{k}_{\max ,\text{D2D}},\:\:\:\sum_{n=1}^{N}\psi _{m}^{n}\hat{p}_{m}^{n}\leq p^{m}_{\max ,\text{Cellular}},  \label{m3455} \\
			& \varphi _{k}^{n}\in \{0,1\},\text{\ }\forall k,n,\text{\ \ \ \ \ \ }\psi_{m}^{n}\in \{0,1\},\text{\ }\forall m,n,\text{{}}  \label{js11}\\
			& \sum_{n=1}^{N}\varphi _{k}^{n}\leq 1,\text{\ }\forall k,\text{\ }\text{ \ \ \ \ \ \ }\sum_{n=1}^{N}\psi _{m}^{n}\leq 1,\text{ \ }\forall m,\text{\ }
			\label{sS11}\\
			&\sum_{k=1}^{K}\varphi _{k}^{n}+\sum_{m=1}^{M}\psi _{m}^{n}\leq L_{\max},\text{ \ }\forall n,\text{\ }
			\label{s11}\\
			&\text{R}_{\text{CU,}m}\geq R^{m}_{\text{CU},\min},~\forall m~~\text{R}_{\text{DU,}k}\geq R^{k}_{\text{DU},\min},~\forall k,	\label{rmin}
			\end{align}}
	\end{subequations}where $\text{R}_{\text{CU,}m}=\sum_{n=1}^{N}\ln (1+\gamma
	_{m,\text{Cellular}}^{n})$ and $\text{R}_{\text{DU,}k}=\sum_{n=1}^{N}\ln (1+\gamma_{k,\text{D2D}}^{n})$. To facilitate the system design, we define 
	$\boldsymbol{\varphi }\in 
	\mathbb{Z}
	^{{\normalsize KN\times 1}}$ and$\ \boldsymbol{\psi }\in 
	\mathbb{Z}
	^{{\normalsize MN\times 1}}$\ as the vectors of subchannel assignment
	variables in\ D2D and cellular networks, respectively. Furthermore, variables {\normalsize $\mathbf{p}$} $\in $ $\mathbb{R}
	^{{\normalsize KN\times 1}}$ and $\ \mathbf{\hat{p}}$ $\in $ $%
	\mathbb{R}
	^{{\normalsize MN\times 1}}$ are the collections of power allocation
	variables in\ D2D and cellular networks, respectively. Variables $%
	P_{\max ,\text{D2D}}$ and $P_{\max ,\text{Cellular}}$\ are the maximum total power for the DUs and CUs,~respectively. $L_{\max}$ denotes the maximum number of CUs that can be paired on
	a subchannel under spectrum sharing scheme\footnote{It can be perceived that at most $(L_{\max}-1)$ CUs share on the same subchannel.}.~Note that the optimization problem (\ref{mm1}) is a
	mixed-integer non-linear programming~(MINLP) because of the interference in
	the rate function, and the presence of the binary constraints. 
	\vspace{-3mm}
	\section{Proposed Solution}
	\vspace{-2mm}
	\textcolor{blue}{One approach to solve a MOOP is the weighted Tchebycheff technique \cite{m3,multiobj}, which offers an auxiliary optimization variable, $\chi $ for (\ref{mm1})}\
	as follows 
	\vspace{-2mm}
	\textcolor{blue}{	\begin{subequations}
		\label{mm6}
		\begin{align}
		& \underset{\left\{ {\normalsize \mathbf{p}},\mathbf{\hat{p}},\boldsymbol{%
				\varphi },\boldsymbol{\psi ,}\chi \right\} }{\text{minimize}}\text{ \ \ \ }\chi  \notag \\
		\mathrm{s.t.}~&(\text{3c})-(\text{3g}), \\
		&\text{ \ }\frac{\alpha}{\text{R}_{\text{DU,max}}} \left( \text{R}_{\text{DU,max}}-\text{R}_{%
			\text{DU}}\right) -\chi \leq 0,  \label{H1} \\
		& \text{ \ }\frac{(1-\alpha)}{\text{R}_{\text{CU,max}}} \left( \text{R}_{\text{CU,max}}-\text{R}_{%
			\text{CU}}\right) -\chi \leq 0,  \label{H2} 
		\end{align}%
\end{subequations}}\textcolor{blue}{where $\alpha$ and $1-\alpha$ are the weighting coefficients\footnote{The non-negative weight $\alpha$ denotes the priorities of CUs and DUs in the resource allocation policy specified in the media access control (MAC) layer to achieve a certain notion of fairness, especially for users who suffer from poor channel conditions.}indicating the impact of the different objectives.} The weighted Tchebycheff method ensures to produce a set of Pareto-optimal solutions when R$_{\text{DU,max}}$ and R$_{\text{CU,max}}$ are the utopian objective points as the maximum of each objective, respectively \cite{multiobj}. In the following, to solve the highly nonconvex optimization problem in (\ref{mm6}) globally, we apply a global optimization approach known as the monotonic optimization method. By
	exploiting monotonicity or hidden monotonicity in the objective function as well
	as constraints, this method guarantees the convergence \cite%
	{mon}. 

	Note that (\ref{mm6}) is not a monotonic optimization problem in canonical
	form since (\ref{H1}) and (\ref{H2}%
	) are not monotonic. To facilitate the presentation, we rewrite these constraints as follows:
	\begin{align}\label{mon1}
	&\sum_{k=1}^{K}\sum_{n=1}^{N}\ln \bigg(1+\frac{\bar{q}^n_k|h_{k}^{n}|^2}{\sigma^{2}_{\text{D2D}}+\sum\limits_{m\in \mathcal{M}}\tilde{p}^n_m|g_{m,k}^{n}|^2+{\textstyle\sum\limits_{j\in \mathcal{K}\backslash k}}\bar{q}^n_{j}|h_{j,k}^{n}|^2}\bigg)\nonumber\\&\geq R_{\text{DU,max}}-\frac{\chi }{\alpha},
	\end{align}
	\begin{align}\label{mon}
	&\sum_{m=1}^{M}\sum_{n=1}^{N}\ln \bigg(1+\frac{\tilde{p}^n_m |g_{m}^{n}|^2}{\sigma^{2}_{\text{CU}}+{\textstyle\sum\limits_{\substack{ i\in\mathcal{M} \backslash m \\g_{i}^{n}<g_{m}^{n}}}}\tilde{p}^n_i|g_{i}^{n}|^2+{\textstyle\sum\limits_{\substack{ j\in \mathcal{K} \\ h_{j}^{n}<g_{m}^{n}}}}\bar{q}^n_j|h_{j}^{n}|^{2}}\bigg)\nonumber\\&\geq R_{\text{CU,max}}-\frac{\chi}{1-\alpha} ,
	\end{align}
	where $\tilde{p}^n_m=\psi _{m}^{n}\hat{p}_{m}^{n}$ and $\bar{q}^n_k=\varphi_{k}^{n}p_{k}^{n}$.~Note that optimization problem (4a) is not monotonic as a result of constraints (4b) and (4c).~First,~the optimization problem in (4a) can be rewritten as a monotonic optimization problem, and then we adopt the polyblock algorithm \cite{mon} to obtain a globally optimal solution.~To doing so,~let us define ${\normalsize \tilde{{\mathbf{{p}}}}_\text{Max}}=\{p^{k}_{\max ,\text{D2D}}\},~\forall k$ and ${\normalsize \tilde{{\mathbf{{q}}}}_\text{Max}}=\{q^{m}_{\max ,\text{Cellular}}\},~\forall m$ as the maximum transmit power for each user over all subchannels. Moreover, the relations in (\ref{mon1}), (\ref{mon}), and (\ref{rmin}) can be equivalently stated as the following single constraints: 
	\begin{align}
	&\underbrace{\sum_{i=1}^{M}\bigg( c_i^{+}({\normalsize \tilde{\mathbf{{p}}}})\bigg)}_{c^{+}(\tilde{{\normalsize {\mathbf{{p}}}}})}-\underbrace{\sum_{i=1}^{M}\bigg(c_i^{-}({\normalsize\tilde{\mathbf{{p}}}})}_{c^{-}(\tilde{{\normalsize {\mathbf{{p}}}}})}\bigg)+R_{\text{DU,max}}-\frac{\chi R_{\text{DU,max}}}{\alpha}
	\geq0  \label{mono1},\\
	&\underbrace{\sum_{j=1}^{K}\bigg(c_j^{+}({\normalsize \bar{\mathbf{{q}}}}
		)\bigg)}_{\hat{c}^{+}(\bar{{\normalsize {\mathbf{{q}}}}})}- \underbrace{\sum_{j=1}^{K}\bigg(c_j^{-}({\normalsize \bar{\mathbf{{q}}}})}_{\hat{c}^{-}(\bar{{\normalsize {\mathbf{{q}}}}})}\bigg)+R_{\text{CU,max}}-\frac{\chi R_{\text{CU,max}}}{1-\alpha}
	\geq 0, \label{mono2}\\
	&\underbrace{\min_{i\in \{1,...,K\}}\left[ c_k^{+}({\normalsize \tilde{\mathbf{{p}}}}) 
		+	\sum\limits_{i\in \mathcal{K}\backslash k}
		c_i^{-}({\normalsize \tilde{\mathbf{{p}}}}) \right]}_{r^{+}({\normalsize \tilde{\mathbf{{p}}}}) } -\underbrace{	\sum\limits_{i\in \mathcal{K}} c_i^{-}({\normalsize \tilde{\mathbf{{p}}}})}_{r^{-}({\normalsize \tilde{\mathbf{{p}}}}) }
	- R^i_{\text{DU,min}}\geq 0,\label{mono3}\\
		&\underbrace{\min_{j\in\{ 1,...,M\}}\left[ c_m^{+}({\normalsize \bar{\mathbf{{q}}}})+	\sum\limits_{j\in \mathcal{M}\backslash m}c_j^{-}({\normalsize \bar{\mathbf{{q}}}}) \right]}_{\hat{r}^{+}({\normalsize \bar{\mathbf{{q}}}})}-\underbrace{	\sum\limits_{j\in \mathcal{M}}c_j^{-}({\normalsize \bar{\mathbf{{q}}}})}_{\hat{r}^{+}({\normalsize \bar{\mathbf{{q}}}})}	
			- R^j_{\text{CU,min}}\geq 0,\label{mono4}
	\end{align}where $c_i^{+}({\normalsize \tilde{\mathbf{{p}}}})$, $c_i^{-}({\normalsize \tilde{\mathbf{{p}}}})$, $c_j^{+}({\normalsize \bar{\mathbf{{q}}}})$, and $c_j^{-}({\normalsize \bar{\mathbf{{q}}}})$ are increasing in ${\normalsize \tilde{\mathbf{ {p}}}}$ and ${\normalsize \bar{\mathbf{ {q}}}}$ given at the top of the next page.
	\begin{figure*}
		\begin{align}
		&c_i^{+}({\normalsize \tilde{\mathbf{{p}}}})=\sum_{n=1}^{N}\ln \bigg(\sigma^{2}_{\text{D2D}}+{\textstyle\sum\limits_{\substack{i\in\mathcal{M} \backslash m \\g_{i}^{n}<g_m^{n}}}}\tilde{p}^n_i|g_{i}^{n}|^2+{\textstyle\sum\limits_{\substack{ j\in \mathcal{K} \\ h_{j}^{n}<g_m^{n}}}}\bar{q}^n_j|h_{j}^{n}|^2+\tilde{p}_m^{n}|g_m^{n}|^2\bigg)\\
		&c_i^{-}({\normalsize \tilde{\mathbf{{p}}}}) =\sum_{n=1}^{N}\ln\bigg(\sigma^{2}_{\text{D2D}}+{\textstyle\sum\limits_{\substack{ i\in\mathcal{M} \backslash m \\g_{i}^{n}<g_m^{n}}}}\tilde{p}^n_i|g_{i}^{n}|^2+{\textstyle\sum\limits_{\substack{ j\in \mathcal{K}, \\ h_{j}^{n}<g_m^{n}}}}\bar{q}^n_j|h_{j}^{n}|^2\bigg),~	c_j^{-}({\normalsize \bar{\mathbf{{q}}}}) =\sum_{n=1}^{N}\ln \bigg(\sigma^{2}_{\text{CU}}+\sum\limits_{m\in \mathcal{M}}\tilde{p}^n_m|g_m^{n}|^2+{\textstyle\sum\limits_{j\in \mathcal{K}\backslash k}}\bar{q}^n_j|h_{j,k}^{n}|^2\bigg),\\
		&c_j^{+}({\normalsize \bar{\mathbf{{q}}}}) =\sum_{n=1}^{N}\ln \bigg(\sigma^{2}_{\text{CU}}+\sum\limits_{m\in \mathcal{M}}\tilde{p}^n_m|g_m^{n}|^2+{\textstyle\sum\limits_{j\in \mathcal{K}\backslash k }}\bar{q}^n_j|h_{j,k}^{n}|^2+\bar{q}^n_k|h_k^{n}|^2\bigg),
		\end{align}
		\hrule
	\end{figure*}
	Bear in mind that equations (\ref{mono1}) and (\ref{mono2}) are the difference
	of two increasing functions.~However, the constraint (\ref{js11}) is a binary constraint which is intractable.~To tackle it, we rewrite (\ref{js11}) in the equivalent form as:
	\vspace{-5mm}
	\begin{align}
	&0\leq\varphi _{k}^{n}\leq 1,~\sum_{k=1}^{K}\sum_{n=1}^{N}\varphi _{k}^{n}-(\varphi _{k}^{n})^{2}\leq 0,\label{R1}\\
	&0\leq\psi _{m}^{n}\leq 1,~\sum_{m=1}^{M}\sum_{n=1}^{N}\psi _{m}^{n}-(\psi _{m}^{n})^{2}\leq 0.\label{R2}
	\end{align}
	It can be perceived that right hand side of (\ref{R1}) and (\ref{R2}) are non-convex as well as non-monotonic.~To tackle this issue, we introduce two slack variables $\nu$ and $\mu$ and rewrite as:
	\begin{align}
	&\sum_{k=1}^{K}\sum_{n=1}^{N}(\varphi _{k}^{n})^{2}+\nu \geq \mathcal{R}_{1},~\sum_{k=1}^{K}\sum_{n=1}^{N}(\varphi _{k}^{n})+\nu \leq \mathcal{R}_{1},\\
	&\sum_{m=1}^{M}\sum_{n=1}^{N}(\psi _{m}^{n})^{2}+\mu\geq \mathcal{R}_{2},~\sum_{m=1}^{M}\sum_{n=1}^{N}\psi _{m}^{n}+\mu\leq \mathcal{R}_{2}, 
	\end{align}
	where $\mathcal{R}_{1}$ and $\mathcal{R}_{2}$ are constant.~We note that the left hand sides of (\ref{R1}) and (\ref{R2}) are monotonically increasing with respect to $\nu$ and $\mu$, respectively.~Consequently, by introducing the auxiliary variable $s_1$,~$s_2$,~$s_3$,~and $s_4$ the problem in (\ref{mm6}) can be then reformulated as following:
	\begin{subequations}
		\label{monokol1}
		\begin{align}
		&\underset{\left\{{\normalsize\tilde{\mathbf{p}},\bar{\mathbf{q}}},s_1,s_2,s_3,s_4,\chi\right\}}{\text{minimize}}\text{ }\chi \\
		& \mathrm{s.t.}~\ 0\leq s_1\leq c^{-}({\normalsize \tilde{{\mathbf{{p}}}}_\text{Max}})-c^{-}({\normalsize \mathbf{0}}),\label{13b}\\
		&\quad\quad c^{-}({\normalsize \tilde{{\mathbf{{p}}}}})+s_1\leq c^{-}({\normalsize \tilde{{\mathbf{{p}}}}_\text{Max}}),\label{13c}\\
		&\quad\quad c^{+}({\normalsize \tilde{{\mathbf{{p}}}}})+s_1\geq c^{-}({\normalsize \tilde{{\mathbf{{p}}}}_\text{Max}}),\label{13d} \\
		&\quad\quad 0\leq s_2\leq \hat{c}^{-}({\normalsize \bar{{\mathbf{{q}}}}_\text{Max}})-\hat{c}^{-}({\normalsize \mathbf{0}}),\label{13e}\\
		&\quad\quad \hat{c}^{-}({\normalsize \bar{{\mathbf{{q}}}}})+s_2\leq  \hat{c}^{-}({\normalsize \bar{{\mathbf{{q}}}}_\text{Max}}),\label{13f}\\
		&\quad\quad \hat{c}^{+}({\normalsize \bar{{\mathbf{{q}}}}})+s_2\geq \hat{c}^{-}({\normalsize \bar{{\mathbf{{q}}}}_\text{Max}}),\label{13g} \\
		&\quad\quad\tilde{p}^n_m\geq 0, \bar{q}^n_k\geq 0, \forall m,k,n, \label{13h}\\
		&\quad\quad \sum_{n=1}^{N}\bar{q}^n_k\leq P_{\max ,\text{D2D}},\:\:\:\sum_{n=1}^{N}\tilde{p}^n_m\leq P_{\max ,\text{Cellular}}, \label{13i}\\
		&\quad\quad0\leq\varphi _{k}^{n}\leq 1,~0\leq\psi _{m}^{n}\leq 1,\label{16j}\\
		&\quad\quad\sum_{k=1}^{K}\sum_{n=1}^{N}(\varphi _{k}^{n})^{2}+\nu \geq \mathcal{R}_{1},~\sum_{m=1}^{M}\sum_{n=1}^{N}(\psi _{m}^{n})^{2}+\mu\geq \mathcal{R}_{2},\label{16k}\\
		&\quad\quad\sum_{k=1}^{K}\sum_{n=1}^{N}(\varphi _{k}^{n})+\nu \leq \mathcal{R}_{1},~\sum_{m=1}^{M}\sum_{n=1}^{N}\psi _{m}^{n}+\mu\leq \mathcal{R}_{2}, \label{16l}\\
		&\quad\quad \text{(\ref{sS11})-(\ref{s11})}, \label{13j} \\
		&\quad\quad\ 0\leq s_3\leq r^{-}({\normalsize \tilde{{\mathbf{{p}}}}_{\text{Max}}})-r^{-}({\normalsize \mathbf{0}}),\label{13bb}\\
		&\quad\quad r^{-}({\normalsize \tilde{{\mathbf{{p}}}}})+s_3\leq r^{-}({\normalsize \tilde{{\mathbf{{p}}}}_{\text{Max}}}),\label{13cc}\\
		&\quad\quad r^{+}({\normalsize \tilde{{\mathbf{{p}}}}})+s_3\geq r^{-}({\normalsize \tilde{{\mathbf{{p}}}}_{\text{Max}}}),\label{13dd} \\
		&\quad\quad\ 0\leq s_4\leq \hat{r}^{-}({\normalsize \bar{{\mathbf{{q}}}}_{\text{Max}}})-\hat{r}^{-}({\normalsize \mathbf{0}}),\label{13bbb}\\
		&\quad\quad \hat{r}^{-}({\normalsize \bar{{\mathbf{{q}}}}})+s_4\leq \hat{r}^{-}({\normalsize \bar{{\mathbf{{q}}}}_{\text{Max}}}),\label{13ccc}\\
		&\quad\quad \hat{r}^{+}({\normalsize \bar{{\mathbf{{q}}}}})+s_4\geq \hat{r}^{-}({\normalsize \bar{{\mathbf{{q}}}}_{\text{Max}}}).\label{13ddd} 
		\end{align}
	\end{subequations}
	Consequently, the feasible set of problem (\ref{monokol1}) can be expressed as the intersection of  two following sets:
\begin{align}
		&\mathcal{G}=\{{\left\{ {\normalsize \tilde{{\mathbf{p}}},\bar{{\mathbf{q}}}},s_1,s_2,s_3,s_4,\chi\right\}
		}:    \tilde{{\mathbf{p}}}\preceq \tilde{{\mathbf{p}}}_\text{ Max}, \bar{{\mathbf{q}}}\preceq \bar{{\mathbf{q}}}_\text{ Max} ,\text{(\ref{13c})},\nonumber\\&\text{(\ref{13f})},\text{(\ref{13i})},\text{(\ref{16j})},\text{(\ref{16l})},\text{(\ref{13j})},\text{(\ref{13cc})},\text{(\ref{13ccc})} \},\\
		&\mathcal{H}=\{{\left\{ {\normalsize \tilde{{\mathbf{p}}},\bar{{\mathbf{q}}}},s_1,s_2,s_3,s_4,\chi\right\}
		}:    \tilde{{\mathbf{p}}}\succeq \mathbf{0}, \bar{{\mathbf{q}}}\succeq \mathbf{0},\text{(\ref{13d})}\nonumber\\& ,\text{(\ref{13g})},\text{(\ref{13h})},\text{(\ref{16k})},\text{(\ref{13dd})},\text{(\ref{13ddd})}\} ,        
		\end{align}where $\mathcal{G}$ and $\mathcal{H}$ are normal and co-normal sets, respectively, in the hyper-rectangle \cite{mon}
\begin{align}
	&\big[0, c^{-}({\normalsize \tilde{{\mathbf{{p}}}}_\text{Max}})-c^{-}({\normalsize \mathbf{0}})\big]\times\big[0,\hat{c}^{-}({\normalsize \bar{{\mathbf{{q}}}}_\text{Max}})-\hat{c}^{-}({\normalsize \mathbf{0}})\big]\times\big[0,\tilde{{\mathbf{ {p}}}}_\text{Max}\big]\nonumber\\
	&\times\big[0,\bar{{\mathbf{ {q}}}}_\text{Max}\big]\times \big[0, c_i^{-}({\normalsize \tilde{{\mathbf{{p}}}}_{\text{Max},k}})-c_i^{-}({\normalsize \mathbf{0}})\big]\nonumber\\
	&\times\big[0,\hat{c}_j^{-}({\normalsize \bar{{\mathbf{{q}}}}_{\text{Max},m}})-\hat{c}_j^{-}({\normalsize \mathbf{0}})\big]\times\big[0,\tilde{{\mathbf{ {p}}}}_{\text{Max},k}\big]\times\big[0,\bar{{\mathbf{ {q}}}}_{\text{Max},m}\big].
	\end{align}
	Finally, it can be shown that problem (\ref{monokol1}) is a monotonic problem. Thus, the optimal solutions can be found at the upper boundary of the feasible set by applying the polyblock algorithm. However, approaching the upper boundary is not possible since it is not precisely known. Hence, we employ the outer polyblock approximation to construct a poly-block. Next, a new poly-block $\mathbf{\Omega}^{{(l+1)}}$ can be constructed from the old one $\mathbf{\Omega}^{{(l)}}$ based on the cutting a cone from the old one, i.e., $\mathbf{\Omega}^{{(l)}}\backslash \mathcal{K}^+_\mathbf{x}$ where the definition of the cone is given in the following.\vspace{-2mm} \begin{definition}
		Assume that $\mathcal{G}$ is
		a normal set in $\mathcal{R}^n_+$, and $\mathbf{y}\in \mathcal{R}^n_+\backslash\mathcal{G}$.
		If $\hat{\mathbf{x}}\in\partial^+\mathcal{G}$ such that $\hat{\mathbf{x}}<\mathbf{y}$, where $\partial^+\mathcal{G}$ is the upper boundary of $\mathcal{G}$, then the cone 	$\mathcal{K}^+_{\hat{\mathbf{x}}}:=\{\mathbf{R}^n_+\:|\: \mathbf{x}>\hat{\mathbf{x}}\}$ isolates $\mathbf{y}$ strictly from $\mathcal{G}$.
	\end{definition}
\vspace{-2mm}

	Let us define vertex set of polyblock $\boldsymbol{\Omega}$ as $\boldsymbol{\omega}$, then $\boldsymbol{w}^{(\ast)}=\{\mathbf{z}\in\boldsymbol{w}\:|\:\mathbf{z}>\mathbf{x} \}$ is the subset of $\boldsymbol{\Omega}$ that involves all vertices in cone $\mathcal{K}^+_{\mathbf{x}}$. The following equation, $ {\boldsymbol{z}}^{(i)}=\boldsymbol{z}+\big({x}^{(i)}-z^{(i)}\big)\boldsymbol{e}_i,\forall i\in\{1,...,n \}$ can be adopted to obtain each vertex $\mathbf{ z}\in\boldsymbol{\Omega}^*$. Note that ${\boldsymbol{z}}^{(i)}$ is obtained by substituting the $i$-th entry of ${\boldsymbol{z}}$ by the $i$-th entry of $\mathbf{x}$. Furthermore, $\boldsymbol{e}_i$ is a unit vector whose elements are equal to one. Let $\boldsymbol{\pi}^{(l)}$ indicate the vertex of $\boldsymbol{\Omega}^{(l)}$ such that maximizes the objective function
	over $\boldsymbol{\Omega}^{(l)}$. The projection operation is given by $\phi(\boldsymbol{\pi}^{(l)})$ which requires solving a one-dimensional problem as $
	\text{max}\{\lambda>0|\lambda \boldsymbol{\pi}^{(l)} \in\mathcal{G}\}$.
	As a result of the normality of $\mathcal{G}$, $\lambda$ can be attained by the bisection search algorithm illustrated in Algorithm 1. Let $\hat{\mathbf{ {x}}}^{(l)}$ express
	the most trustworthy feasible solution at the $l$-th iteration, and $V^{(l)} = f(\hat{\mathbf{ {x}}}^{(l)})$ be the current most suitable value. At the next iteration $(l+1)$, if $\phi(\boldsymbol{\pi}^{l+1})\in\mathcal{G}\bigcap\mathcal{H}$ and $f(\phi(\boldsymbol{\pi}^{l+1}))\geq V^{(l)}$, we have $\hat{\mathbf{ {x}}}^{l+1}=\phi(\boldsymbol{\pi}^{l+1})$ and $V^{l+1}=f(\phi(\boldsymbol{\pi}^{l+1}))$. Otherwise, let $\hat{\mathbf{ {x}}}^{l+1}=\hat{\mathbf{ {x}}}^{l}$ and $V^{l+1}=V^{l}$. The algorithm stops if $|f(\boldsymbol{\pi}^{(l)})-V^{(l)}|\leq \epsilon$ satisfies, where $\epsilon\geq 0$ is a given tolerance. Moreover, $\hat{\mathbf{ {x}}}^{l}$ is said to be an $\epsilon$-optimal solution if $f({\mathbf{ {x}}}^{\ast})-\epsilon\leq f(\hat{\mathbf{ {x}}}^{l+1})\leq f({\mathbf{ {x}}}^{\ast})$. We summarize the polyblock approach in Algorithm 2. 
	\begin{algorithm}[t]
		\caption{Bisection Projection Search Algorithm}
		\begin{algorithmic}[1]
			\renewcommand{\algorithmicrequire}{\textbf{Input:}}
			\renewcommand{\algorithmicensure}{\textbf{Output:}}
			\REQUIRE $\boldsymbol{\pi}^{(l)}$ and $\mathcal{G}$\\
			\textbf{Output} $\lambda$ such that $\lambda=\text{ argmax}\{\lambda>0\:|\: \lambda \boldsymbol{\pi}^{(l)} \in \mathcal	{G} \}$
			\STATE Set $\lambda_\text{min}=0$, $\lambda_\text{max}=1$, and the error tolerance $\delta\ll1$.\\
			\STATE \textbf{repeat}\\
			\STATE \quad Let ${\bar{\lambda}}=\frac{\lambda_\text{min}+\lambda_\text{max}}{2}$
			\STATE \quad Solve the feasibility problem $(\ref{monnn})$.
			\STATE \quad\quad \textbf{If}
			\STATE \quad\quad\quad Check if $\bar{\lambda}$ is feasible, i.e., $\bar{\lambda} \boldsymbol{\pi}^{(l)}\in \mathcal	{G}$, then set\\ \quad\quad\quad $\lambda_\text{min}={\bar{\lambda}}$
			\STATE \quad\quad \textbf{else}
			\STATE \quad\quad\quad Set $\lambda_\text{max}={\bar{\lambda}}$		
			\STATE   \textbf{until}\: $\lambda_\text{max}-\lambda_\text{min}<\delta$. 
			\STATE   \textbf{Return} ${\lambda}=\lambda_\text{min}$.
		\end{algorithmic}
	\end{algorithm}Taking into account that the projection of $\boldsymbol{\pi}^{(l)}$ in the $l$-th iteration over set $\mathcal{G}$ is required. Therefore, we can obtain the projection by $\phi(\boldsymbol{\pi}^{(l)})=\lambda\boldsymbol{\pi}^{(l)}$, where $\lambda$ is the projection parameter. Moreover, ${\lambda}$ is given by ${\lambda}=\text{max}\{\alpha\:|\:\alpha\boldsymbol{\pi}^{(l)}\in \mathcal{G}\}$, where ${\lambda}\in [0,1]$. In particular, the bisection search technique can be applied to obtain ${\lambda}$. For a given $\lambda$ and vertex $\boldsymbol{\pi}^{(l)}$ in the $l$-th iteration, the following feasibility problem needs to be solved:
	\begin{subequations}
		\label{monnn}
		\begin{align}
		&\max_{\textbf{y}} ~ 1\label{monn-12}
		\\
		\text{ s.t.} &~\mathbf{y}\in \mathcal{G}.
		\end{align}
	\end{subequations}
	We can obtain the optimal value for subcarrier allocations ($\phi^{n}_{k}$, $\psi^{n}_{m}$) via comparing the values of the entries of the power allocations ($\tilde{{\mathbf{ {p}}}}$,~$\tilde{{\mathbf{{q}}}}$) with zero. If the values of ($\tilde{{\mathbf{ {p}}}}$,~$\tilde{{\mathbf{{q}}}}$)  be greater than 0, it means that the corresponding subcarrier allocations ($\phi^{n}_{k}$, $\psi^{n}_{m}$) would be zero. The details of the projection bisection search algorithm are given in Algorithm 2.
	\vspace{-5mm}
 \section{Computational Complexity}
			\vspace{-2mm}
				The polyblock algorithm is influenced by the configuration of the objective function and the constraints that create the normal set. First, the most suitable vertex by its projection on the normal set is determined. Then, the projection of the picked vertex is obtained. Eventually, we obtained the new vertex set by eliminating the inappropriate vertices. In particular, the dimension of problem (P1), the number of iterations required for convergence, and the number of iterations expected for the projection of each vertex are assumed to be $B_{1}$, $B_{2}$ and $B_{3}$, respectively. In summary, the complexity order can be stated as $\mathcal{O}(B_{2}(B_{2}\times B_{1}+B_{3}))$ \cite{mon,Ata_Mon}.
	\begin{algorithm}[t]
		\caption{Outer Poly-block Approximation Algorithm}
		\begin{algorithmic}[1]
			\renewcommand{\algorithmicrequire}{\textbf{Input:}}
			\renewcommand{\algorithmicensure}{\textbf{Output:}}
			\REQUIRE An function $f(\cdot)$, a compact normal set $\mathcal{G}$ and a closed conormal set $\mathcal{H}$, such that $\mathcal{G}\bigcap\mathcal{H}\neq {\O}$
			\STATE \textbf{Output}: an $\epsilon$-optimal solution $\mathbf{x}^*$
			\STATE Initialize iteration index $l=0$, poly-block $\mathbf{\Omega}^{(1)}$ be box $[0,\mathbf{b}]$ that encloses $\mathcal{G}\bigcap\mathcal{H}$ with vertex set $\boldsymbol{\omega}^{(1)}=\{\mathbf{b}\}$. $\epsilon$ is a small positive number. $V^{(l)}=-\infty$ 
			\STATE \textbf{Repeat}\\
			\STATE \quad From $\boldsymbol{\omega}^{(l)}$, select, $\boldsymbol{\pi}^{(l)}\in \text{argmax}\{f(\boldsymbol{\pi})|\boldsymbol{\pi}\in \boldsymbol{\omega}^{(l)}\}$ .
			\STATE \quad obtain the projection of $\boldsymbol{\pi}^{(l)}$, i.e., $\phi(\boldsymbol{\pi}^{(l)})$ on the upper \\ \quad boundary of $\mathcal{G}$
			\STATE \quad\quad \textbf{if}  $\phi(\boldsymbol{\pi}^{(l)})=\boldsymbol{\pi}^{(l)}$, i.e., $\boldsymbol{\pi}^{(l)}\in \mathcal{G}$ \textbf{then}
			\STATE \quad\quad\quad $\hat{\mathbf{x}}^{(l)}=\boldsymbol{\pi}^{(l)}$ and $V^{(l)}=f(\boldsymbol{\pi}^{(l)})$
			\STATE \quad\quad \textbf{else} 
			\STATE  \quad\quad\quad if $\phi(\boldsymbol{\pi}^{(l)})	\in\mathcal{G}\bigcap\mathcal{H}\neq {\O}$ and $f(\phi(\boldsymbol{\pi}^{(l)}))\geq V^{l-1}$, \\ \quad\quad\quad then let then best value be $\hat{\mathbf{ {x}}}^{(l)}=\phi(\boldsymbol{\pi}^{(l)})$ and\\ \quad\quad\quad ${\small V^{l}=f(\phi(\boldsymbol{\pi}^{(l)}))}$. Otherwise, ${\small { \hat{\mathbf{ {x}}}^{(l)}=\hat{\mathbf{ {x}}}^{l-1}}}$ and\\ \quad\quad\quad ${\small V^{l}=V^{l-1}}$.
			\STATE \quad\quad\quad Let ${\mathbf{ {x}}}=\phi(\boldsymbol{\pi}^{(l)})$ and 
			\\ \quad\quad\quad $ {\small \boldsymbol{w}^{(l+1)}=(\boldsymbol{w}^{(l)}\backslash  \boldsymbol{w}^{(\ast)})\bigcup \big\{{\boldsymbol{z}}^{(i)}=\boldsymbol{z}+\big({x}^{(i)}-z^{(i)}\big)\boldsymbol{e}_i|}$\\ \quad\quad\quad $\boldsymbol{z}\in \boldsymbol{w}^{(\ast)},\forall i\in\{1,...,n \}\big\}$, where $\boldsymbol{w}^{(\ast)}=\{\mathbf{z}\in$\\ \quad\quad\quad $\boldsymbol{w}^{l}\:|\:\mathbf{z}>\mathbf{x} \}$.
			\STATE \quad\quad\quad Remove from $\boldsymbol{w}^{(l+1)}$ the improper vertices and the\\ \quad \quad\quad vertices $\{\mathbf{z}\in \boldsymbol{w}^{l+1}\:|\:\mathbf{z} 
			\notin
			\mathcal{H}\}$.
			\STATE \quad\quad \textbf{end} 
			\STATE \quad  Set $l$=$l$+1.\\ \vspace{1mm}
			\STATE   \textbf{Until} $|f(\boldsymbol{\pi}^{(l)})-V^{(l)}|\leq \epsilon$.\\\vspace{1mm}
			\STATE   \textbf{Return} $\mathbf{x}^\ast=\hat{\mathbf{x}}^{(l)}$.
		\end{algorithmic}
	\end{algorithm}

	\vspace{-4mm}
	\section{Simulation Results}
		\vspace{-2mm}
In this section, we evaluate the performance of the proposed algorithm by numerical simulation. For each subchannel, a Rayleigh flat fading involving the path-loss model is considered.
	The simulation parameters are given in Table \ref{Simulation Parameters}, unless otherwise specified.\begin{table}[t]
		\centering
		\caption{simulation parameters}
		\label{Simulation Parameters}
		\begin{tabular}{|c|c|}\hline
			{\bf Parameter} & {\bf Value} \\ \hline \hline
			{Cell diameter} & {$250$ m}\\ \hline
			{Distance between D2D link}&{$30$m}\\\hline
			{Number of CUs ($M$)} & {$6$} \\\hline
			{Number of DUs  ($K$)} & {$3$} \\\hline
			{Number of sub-carriers ($N$)} & {$4$} \\\hline
			{Noise power} & {$-120$} dBm \\\hline
			Sub-carrier bandwidth & {$180$} kHz \\\hline
			{Path-loss model for cellular links} & {$128.1+37.6\log(d)$} \\\hline
			{Path-loss model for D2D links} & {$148.1+40\log(d)$} \\\hline
			Maximum transmit power of the DUs & {$25$ dBm} \\\hline
			{Minimum data rate requirement ($\overline{R}_\textnormal{min}$)} & {$1$ bps$/$Hz} \\\hline
		\end{tabular}
	\end{table}
	\begin{figure}\label{Noma}
		\centering
		\includegraphics[width=2.4in] {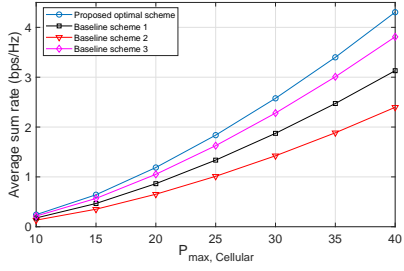}
		\caption{\textcolor{blue}{Average sum rate versus maximum transmit power of CUs.}}
	\end{figure}
 	\begin{figure}
	\centering
	\includegraphics[width=2.2in] {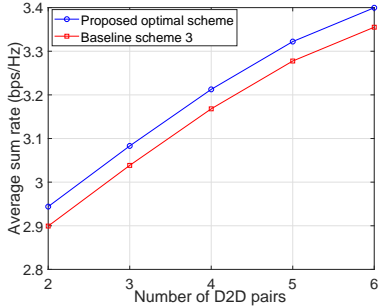}
	\caption{\textcolor{blue}{Average sum rate versus maximum number of D2D pairs.}}
\end{figure}

\textcolor{blue}{Fig.~1 shows the system sum-rate versus maximum transmit power of CUs for an equal weight of CUs and DUs i.e., $\alpha=0.5$.~It can be seen that by increasing $p_{\max,\text{Cellular}}$, not only the CUs more easily meet their minimum requirement but also the system throughput is improved. In fact with the increase $p_{\max,\text{Cellular}}$, the system throughput increases monotonically due to enhancing the achievable rate for CUs which results in an enhancement of the total throughput of the system.~This figure also shows that large value of $p_{\max,\text{Cellular}}$ leads to releasing the potential of DUs that enables more D2D pairs to access the network.}~\textcolor{blue}{For comparison, we also consider three baseline schemes for the system sum rate.~For baseline scheme 1, we adopt a subcarrier assignment randomly where the power allocation is obtained based on our proposed monotonic approach.~For baseline scheme 2, we consider MC-OMA in which each subcarrier is allocated to at most one user.}~\textcolor{blue}{For baseline scheme 3, we considered the proposed method in [12] in which an iterative algorithm is adopted to find the resource allocation policy.~It can be perceived that our proposed algorithm outperforms the proposed algorithm in [12] due to performing a monotonic approach which gives us optimal resource allocation policy. Furthermore, we observe that the system sum rate for MC-OMA achieves a lower system sum rate compared to MC-NOMA due to underutilizing the orthogonal subcarrier assignment.}
	
	\textcolor{blue}{Fig.~2 illustrates the sum-rate versus the number of D2D pairs.~It can be perceived  that our proposed solution significantly ameliorates the system sum rate as compared to baseline scheme 3.~This is due to the fact that in our proposed algorithm the spectrum resources are profoundly reused and our proposed algorithm outperforms the baseline scheme 3 with SIC scheme.}
	
	 \textcolor{blue}{The trade-off between the total data rate of DUs and CUs is investigated in Fig. 3.
	This figure is obtained by solving problem (18) for different values of $\alpha$ $\in [0,1]$, with a step size of $0.1$.
	It can be seen that a non-trivial trade-off between the total data rate of DUs and CUs exists.
	In particular,~the total DUs data rate is a decreasing function versus the total CUs data rate.In other words, maximizing the total DUs data rate leads to a reduction in total CUs data rate due to conflicting objectives functions.~\textcolor{blue}{In fact, by changing the weight factor, we can provide fairness between the cellular and DUs.}
	This figure also demonstrates the superiority of the NOMA scheme as compared to the conventional OMA method.} 
		\vspace{-5mm}
	\section{Conclusion}
		\vspace{-1mm}
	\textcolor{blue}{In this paper, we investigated the tradeoff between DUs and CUs in
	uplink underlaying CUs-enabled NOMA networks.~We formulated a MOOP framework which jointly maximizes the throughput of DUs and CUs simultaneously, to obtain power allocation strategy and subchannel assignment. The MOOP was converted into a SOOP using a weighted Tchebychef method and then solved via monotonic optimization to obtain an optimal solution.~Simulation results not only unveiled an interesting tradeoff between
	the studied competing objective functions but also investigated the superiority of our proposed scheme as compared to MC-OMA.}

	\begin{figure}[t]
	\centering
	\includegraphics[width=8.00cm,height=3.700cm] {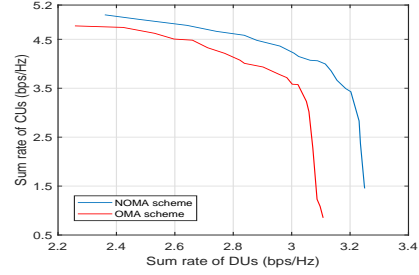}
	\caption{\textcolor{blue}{Throughput trade-off region of CUs and DUs.}}\label{Region}
\end{figure}

\end{document}